\documentclass[aps,prb,twocolumn,showpacs,superscriptaddress,floatfix]{revtex4}
\usepackage{graphicx}

\begin{document}

\author{S.\ Miyahara}
\affiliation{Department of Physics, Aoyama Gakuin University, Sagamihara,
Kanagawa 229-8558, Japan}
\author{J.-B.\ Fouet}
\affiliation{Institut Romand de Recherche Num\'erique en Physique
des Mat\'eriaux (IRRMA), PPH-Ecublens, CH-1015 Lausanne,
Switzerland}
\author{S.R.\ Manmana}
\affiliation{Institut f\"ur Theoretische Physik III, Universit\"at
  Stuttgart, Pfaffenwaldring 57, D-70550 Stuttgart, Germany}
\affiliation{Fachbereich Physik, Philipps Universit\"at Marburg,
D-35032 Marburg, Germany}
\author{R.M.\ Noack}
\affiliation{Fachbereich Physik, Philipps Universit\"at Marburg,
D-35032 Marburg, Germany}
\author{H. Mayaffre}
\affiliation{Laboratoire de Spectrom\'etrie Physique,
Universit\'e J. Fourier \& UMR5588 CNRS, BP 87, 38402, Saint
Martin d'H\`{e}res, France}
\author{I. Sheikin}
\affiliation{Grenoble High Magnetic Field Laboratory, CNRS,
BP 166, F-38042 Grenoble Cedex 09, France}
\author{C. Berthier}
\affiliation{Laboratoire de Spectrom\'etrie Physique,
Universit\'e J. Fourier \& UMR5588 CNRS, BP 87, 38402, Saint
Martin d'H\`{e}res, France}
\affiliation{Grenoble High Magnetic Field Laboratory, CNRS,
BP 166, F-38042 Grenoble Cedex 09, France}
\author{F.\ Mila}
\affiliation{Institute of Theoretical Physics, \'Ecole Polytechnique
F\'ed\'erale de Lausanne, CH-1015 Lausanne, Switzerland}

\title{Uniform and staggered magnetizations induced by
  Dzyaloshinskii-Moriya interactions in isolated and coupled spin 1/2
  dimers in a magnetic field}

\date{\today}

\begin{abstract}
We investigate the interplay of Dzyaloshinskii-Moriya interactions and
an external field in spin 1/2 dimers.  For isolated dimers and at low
field, we derive simple expressions for the staggered and uniform
magnetizations which show that the orientation of the uniform
magnetization can deviate significantly from that of the external
field.  In fact, in the limit where the ${\bf D}$ vector of the
Dzyaloshinskii-Moriya interaction is parallel to the external field,
the uniform magnetization actually becomes {\it perpendicular} to the
field.  For larger fields, we show that the staggered magnetization of
an isolated dimer has a maximum close to one-half the polarization,
with a large maximal value of $0.35\,g\mu_B$ in the limit of very
small Dzyaloshinskii-Moriya interaction.  We investigate the effect of
inter-dimer coupling in the context of ladders with Density Matrix
Renormalization Group (DMRG) calculations and show that, as long as
the values of the Dzyaloshinskii-Moriya and of the exchange
interaction are compatible with respect to the development of a
staggered magnetization, the simple picture that emerges for isolated
dimers is also valid for weakly coupled dimers with minor
modifications. The results are compared with torque measurements on
Cu$_{2}$(C$_{5}$H$_{12}$N$_{2}$)$_{2}$Cl$_{4}$.
\end{abstract}


\maketitle

\section{Introduction}
In Mott insulators, the Heisenberg interaction $J {\bf S}_i.{\bf S}_j$
is in most cases the dominant source of coupling between local
moments, and most theoretical investigations are based on modeling in
which only this type of interaction is included.  It has been known
for a very long time, however, that other, less symmetric,
interactions are present.  For instance, unless there is an inversion
center on a bond, spin-orbit coupling induces an antisymmetric
interaction of the form ${\bf D}\cdot({\bf S}_i \times {\bf S}_j)$,
which is known as the Dzyaloshinskii-Moriya (DM)
interaction.\cite{Dzyaloshinskii58,Moriya60} Since it breaks the
fundamental SU(2) symmetry of the Heisenberg interactions, the DM
interaction is at the origin of many deviations from pure Heisenberg
behavior, such as canting\cite{Coffey90} or small
gaps.\cite{Sakai94,Dender97,Oshikawa97,Zhao03,Fouet04,Chernyshev05,Fouet06}
It is also known to have a dramatic impact on the properties of
antiferromagnets in a magnetic field.  Numerous experimental
investigations of quantum antiferromagnets currently in progress in
large field facilities call for a detailed understanding of this
problem.\cite{Chaboussant98,Kageyama99,Cepas01,Coldea02,Jaime04}
Several issues have recently been the subject of rather intensive
research.  For instance, the impact on triplon Bose-Einstein
condensation\cite{Affleck90,Giamarchi99,Matsumoto02} of DM
interactions has been analyzed.\cite{Sirker04} The interplay of
frustration and DM interactions has also received significant
attention.\cite{Elhajal02,Elhajal05,Kotov05} The consequence of the
breaking of SU(2) symmetry on the excitation spectrum is also well
understood thanks to the work of several people including some of the
present
authors.\cite{Sakai94,Dender97,Oshikawa97,Zhao03,Fouet04,Chernyshev05,Fouet06}
It is by now well established that a DM interaction can open a gap in
otherwise gapless regions.  The scaling of this gap with the magnitude
of the DM interaction has been worked out for several
cases.\cite{Oshikawa97,Fouet04}

Surprisingly, however, the other important consequence of the breaking
of the SU(2) symmetry on the ground state properties of weakly coupled
dimers, namely the development of a local magnetization, has not
received much attention so far, although it is of immediate relevance
to several compounds.  It was shown in the case of
SrCu$_2$(BO$_3$)$_2$ that a DM interaction can lead to the development
of a measurable (and in fact quite large) staggered
magnetization,\cite{Kodama05} but a simple picture of how the
magnitude and the orientation of the DM interaction with respect to
the magnetic field influences these properties has not yet emerged.
Besides, the fact that a DM interaction can lead to the development of
a transverse uniform magnetization and its impact on torque
measurements of the magnetization have not been investigated in
detail.  All these questions are central to the understanding of
several systems of current interest.  In particular, recent NMR
results by Cl\'emancey {\em et al.}\cite{Clemancey06} have revealed
the presence of a staggered magnetization in the dimer compound
Cu$_{2}$(C$_{5}$H$_{12}$N$_{2}$)$_{2}$Cl$_{4}$ [abbreviation: Cu(Hp)Cl],
and the interpretation of these results requires a precise
investigation of the effect of DM interactions on weakly coupled dimer
systems.

In this paper, we present a systematic analysis of the consequences of
an intra-dimer DM interaction on the development of local
magnetization in systems of weakly coupled spin 1/2 dimers.  We first
look at the case of an isolated dimer, and derive simple expressions
in the limits of weak and strong magnetic field which we believe are
very useful to get a simple picture of subtle effects such as the
effect of the relative orientation of the magnetic field and the ${\bf
D}$ vector of the DM interaction on the uniform magnetization.  We
then turn to the case of coupled dimers and concentrate on a simple
ladder geometry.  This choice is motivated partly by the potential
relevance of this geometry to actual compounds such as Cu(Hp)Cl, and
by the possibility to obtain very accurate results using the Density
Matrix Renormalization Group method
(DMRG)\cite{white,schollwoeck05,noack05} in this quasi-one dimensional
geometry. Finally, we report new torque measurements on Cu(Hp)Cl and
discuss them in the light of these results.

\section{Isolated Dimer}

The problem of an isolated dimer in a magnetic field in the presence
of a DM interaction is defined by the Hamiltonian
\begin{equation}
  H  =  J{\bf S}_{1} \cdot {\bf S}_{2}
+ {\bf D} \cdot ( {\bf S}_{1} \times {\bf S}_{2})- g \mu_B H (S_{1}^z+S_{2}^z).
\label{eq-ham-dimer}
\end{equation}

\begin{figure}[bht]
\begin{center}
  \includegraphics[width=\columnwidth]{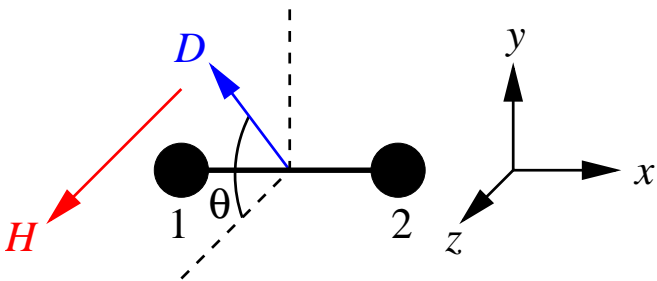} \\
\end{center}
\caption{Pictorial representation of the model of
  Eq.~(\protect{\ref{eq-ham-dimer}}) of a dimer with DM interaction in
  a magnetic field.}
\label{fig:model-dimer}
\end{figure}

The $z$ axis has been chosen to be that of the magnetic field, and the
$yz$ plane as the plane defined by the magnetic field and the ${\bf D}$
vector (see Fig.~\ref{fig:model-dimer}).
In actual systems, the direction of the ${\bf D}$ vector is fixed by
the microscopic arrangement of atoms and orbitals, and it is the
orientation of the magnetic field that can be varied with respect to the
crystal, but the
convention of having the magnetic field along the $z$ axis makes the
discussion somewhat simpler.
The ${\bf D}$ vector is written as ${\bf D}=(0,D \sin \theta, D \cos
\theta)$.

The isolated dimer problem is, of course, very simple.  The Hilbert
space is of dimension 4, and it will prove convenient to work in the
basis
\begin{eqnarray}
  | s \rangle & = & \frac{1}{\sqrt{2}} \left( | \uparrow \downarrow \rangle
  - | \downarrow \uparrow \rangle \right), \nonumber \\
  | t_1 \rangle & = & | \uparrow \uparrow \rangle, \nonumber \\
  | t_0 \rangle & = & \frac{1}{\sqrt{2}} \left( | \uparrow \downarrow \rangle
  + | \downarrow \uparrow \rangle \right), \nonumber \\
  | t_{-1} \rangle & = & | \downarrow \downarrow \rangle.
\end{eqnarray}
Unfortunately, the explicit expressions that
can be derived for the eigenenergies and eigenvectors are cumbersome
and not particularly useful.
Therefore, we will discuss the results from the point of view of
symmetry, will derive useful expressions for small ${\bf D}$ in weak
field and close to saturation, and will present plots of some
representative exact results.

\subsection{Symmetry analysis}

In a magnetic field, the SU(2) symmetry of the Heisenberg model is
reduced to a U(1) symmetry corresponding to a rotation around the
field direction.

As soon as a DM interaction with a ${\bf D}$ vector not parallel to
the magnetic field is introduced, the rotational symmetry in spin
space is completely lost.
The only symmetry that remains is the mirror symmetry with respect to
the $yz$ plane (the plane containing the magnetic field and the
${\bf D}$ vector), which exchanges sites 1 and 2 and simultaneously
changes the sign of the $x$-component of the spin operators (the
component perpendicular to the mirror plane).
As a consequence, the expectation values of local spin operators in
any eigenstate of the Hamiltonian satisfy the relations:
\begin{eqnarray}
\langle S_{1}^{x}\rangle & = &  -\langle S_{2}^{x}\rangle \nonumber \\
\langle S_{1}^{y}\rangle & = & \langle S_{2}^{y}\rangle \nonumber \\
\langle S_{1}^{z}\rangle & = & \langle S_{2}^{z}\rangle \; .
\label{eq-sym}
\end{eqnarray}
These relations imply that the staggered magnetization per site,
defined as ${\bf m}_s=(\langle {\bf S}_{1}-{\bf S}_2 \rangle)/2$, is
perpendicular to the plane defined by the magnetic field and the ${\bf
  D}$ vector, while the uniform magnetization per site defined by
${\bf m}_u=(\langle {\bf S}_{1}+{\bf S}_2 \rangle)/2$ must lie in that
plane.

If the ${\bf D}$ vector is parallel to the field, the U(1) rotational
symmetry is still present, which can be easily checked since
$S_z^{tot}=S_{1}^{z}+S_{2}^{z}$ commutes with
$S_{1}^{x}S_{2}^{y}-S_{1}^{y}S_{2}^{x}$.  The states $\vert
t_{-1}\rangle$ and $\vert t_{1}\rangle$ are still eigenstates with
energies $J/4\pm g\mu_BH$, but $\vert s \rangle$ and $\vert
t_0\rangle$ get coupled.  The staggered magnetization is identically
zero, while the uniform magnetization jumps abruptly from 0 to
$2g\mu_B \hat z$ ($ \hat z$ is the direction of the applied magnetic
field) at a critical field $H_c$ larger than its $D=0$ value
$J/g\mu_B$.

\subsection{Low-field limit}

In the limit $D/J\ll 1$ and below the saturation field $H_c=J/g\mu_B$,
the ground-state wave function, up to second order in $D/J$, reads:
\begin{eqnarray}
  |\phi_0 \rangle & = & \left(1 - \frac{D^2}{4 J^2} \right) |s \rangle
  - \frac{D \sin\theta}{2 \sqrt{2} (J - g \mu_B H) } |t_1 \rangle \nonumber \\
  & + & i \frac{D \cos\theta}{2 J} |t_0 \rangle
  - \frac{D \sin\theta}{2 \sqrt{2} (J + g \mu_B H)} |t_{-1} \rangle.
\end{eqnarray}

In the low-field limit, first-order perturbation theory in $H$ can be
used to derive simple expressions for the expectation value of the
various spin operators:
\begin{eqnarray}
  \langle S_1^x \rangle & = &
  -\langle S_2^x \rangle = \frac{g \mu_B H D \sin\theta}{2 J^2}, \nonumber\\
  \langle S_1^y \rangle & = &
  \langle S_2^y \rangle
  = -\frac{g \mu_B H D^2 \cos\theta \sin\theta}{4 J^3} , \nonumber \\
  \langle S_1^z \rangle & = &
  \langle S_2^z \rangle
  = \frac{g \mu_B H D^2 \sin^2\theta}{4 J^3}.
\end{eqnarray}
These expressions lead to compact and suggestive expressions for the
uniform and staggered magnetizations:
\begin{eqnarray}
  {\bf m}_u & = &
  \frac{g \mu_B}{4 J^3} \left({\bf D} \times \bf H \right) \times {\bf D}, \nonumber \\
  {\bf m}_s & = &
  \frac{g \mu_B}{2 J^2} \left( {\bf D} \times {\bf H} \right).
  \label{eq-m-Hsmall}
\end{eqnarray}

As required by symmetry, the staggered magnetization is perpendicular
to both the field and the ${\bf D}$ vector.  As far as the uniform
magnetization is concerned, symmetry only requires that it lies in the
plane of the magnetic field and of the ${\bf D}$ vector, but in the
low field limit, Eq.~(\ref{eq-m-Hsmall}) shows that it is perpendicular
to the ${\bf D}$ vector.  So the uniform magnetization is in general
{\it not} parallel to the magnetic field, as it would be in a system
with SU(2) symmetry, and it can in fact deviate strongly: In the limit
where the ${\bf D}$ vector becomes parallel to the field, the uniform
magnetization becomes perpendicular to the magnetic field, a rather
anomalous behavior that should have important consequences for torque
measurements of the magnetization.

Another remarkable feature of these results is that the staggered
magnetization is first order in ${\bf D}$, while the uniform
magnetization is second order.
Thus, at low field the response is dominated by the staggered
magnetization, as already observed in SrCu(BO$_3$)$_2$.

Finally, let us emphasize that, as implied by Eq.~\ref{eq-m-Hsmall},
the uniform and staggered magnetizations have universal expressions
in terms of the magnetic field and of the ${\bf D}$ vector, which are valid
regardless of their orientation with respect to the lattice.

\subsection{Critical field}

At the critical field $H_c=J/g\mu_B$, one has to turn to degenerate
perturbation theory since, for $D=0$, $\vert s \rangle$ and $\vert
t_{1}\rangle$ are degenerate.  When the ${\bf D}$ vector is not
parallel to the field, these states get coupled by an off-diagonal
term $D \sin \theta$.  The ground state wave function is then simply
given by $\phi_0=(\vert s \rangle - \vert t_1 \rangle)/\sqrt{2}$, and
the staggered magnetization per site is equal to
$(\sqrt{2}/4)g\mu_B\simeq 0.35\,g\mu_B$.  Interestingly enough, this
maximal value is independent of the angle $\theta$ and does not vanish
but remains quite large in the limit where $D$ goes to zero.  Note,
however, that the staggered magnetization only takes significant values
close to $H=J/g\mu_B$, in an interval of width of the order of $D\sin
\theta$ which shrinks to zero in the limit where $D$ goes to zero,
consistent with a vanishing staggered magnetization when $D=0$.

When $D \ll J$, the uniform magnetization at this field is equal to
$g\mu_B$, which corresponds to half the polarization value.  When the
angle between $D$ and $H$ is not $\pi/2$, a small uniform component
develops along $y$ due to the coupling of $\vert s\rangle$ with $\vert
t_0 \rangle$. This transverse (with respect to the field) uniform
magnetization is given by $m_u^y=-(\sqrt{2}/4)\cos \theta (D/J) g\mu_B
\simeq -0.35 \cos \theta (D/J) g\mu_B$.  In contrast to the small
field result, it is now linear in $D$, but remains much smaller than
the staggered magnetization, which is of order one.

\subsection{Exact results}

\begin{figure}
\begin{center}
  \includegraphics[width=\columnwidth]{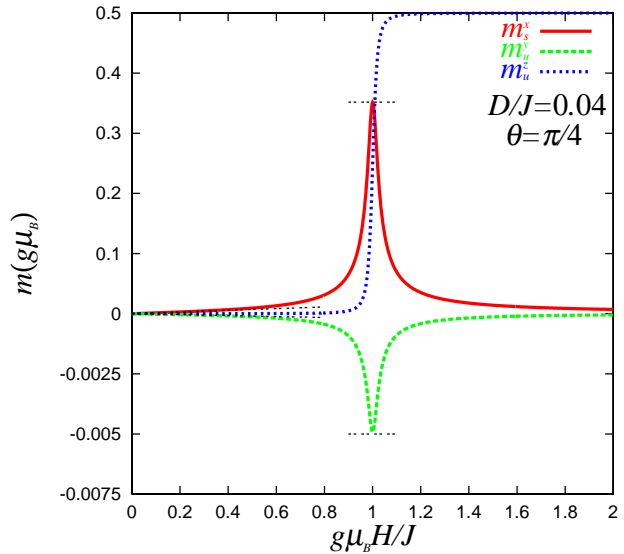}
\end{center}
\caption{Field dependence of the uniform and staggered magnetizations
per site $m_u^y$, $m_u^z$ and $m_s^x$ for $D/J = 0.04$ and $\theta = \pi/4$.
The dashed lines are the analytical results derived in the limit $D/J\ll 1$. Note the difference
of scale for positive and negative magnetizations.}
\label{fig-D0.04}
\end{figure}

To get an idea of the accuracy of the expressions obtained at low
field and close to the saturation field, we have plotted in
Fig.~\ref{fig-D0.04} the exact value of $m_s^x$, $m_u^y$ and $m_u^z$
for a representative case ($D/J=0.04$ and $\theta=\pi/4$).
The small field expression is quantitatively accurate up to
$H\simeq0.25 J/g\mu_B$, and the width of the peak of the staggered
magnetization and the maximal value of $m_u^y$ are indeed of order
$D$.

\section{Coupled dimers (ladder)}

\subsection{The model}

\begin{figure}[bht]
\begin{center}
  \includegraphics[width=.8\columnwidth]{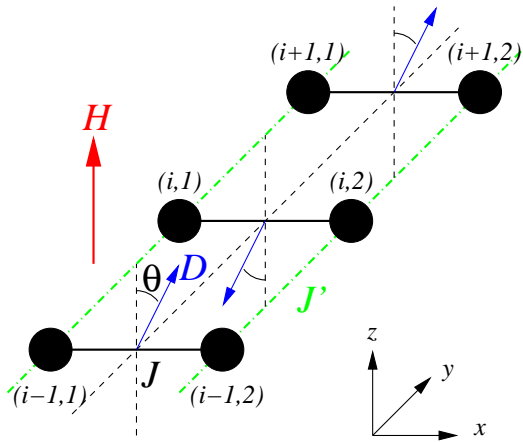}
\end{center}
\caption{Ladder with staggered DM interaction.}
\label{fig:model_ladder}
\end{figure}

\begin{figure}[bht]
\begin{center}
  \includegraphics[width=\columnwidth]{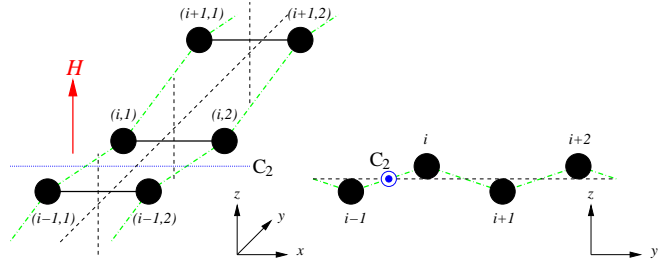}
\end{center}
\caption{Structure of a buckled ladder. In such a ladder,
 a staggered DM interaction in the $y$-direction is allowed by symmetry.}
\label{fig:model_zigzag}
\end{figure}

In this section, our goal is to check to which extent the properties
of a system of weakly coupled dimers resemble those of isolated
dimers.  In particular, the transition between zero magnetization and
polarization takes place through an extended region of magnetic field
of the order of the inter-dimer coupling, and we would like to know
how the system behaves within and outside this region.  We will attack
this problem numerically, and in order to perform simulations on large
systems, we have chosen to work in a ladder geometry and to use the
DMRG.  The model is defined by the Hamiltonian
 \begin{eqnarray}
  H & = & J \sum_{i} {\bf S}_{i,1} \cdot {\bf S}_{i,2}
  +  \sum_{i} (-1)^i \ {\bf D} \cdot ( {\bf S}_{i,1} \times {\bf S}_{i,2})\nonumber \\
  & + & J_\parallel \sum_{i} ({\bf S}_{i,1} \cdot {\bf S}_{i+1,1}+ {\bf S}_{i,2} \cdot {\bf S}_{i+1,2})
 \nonumber \\ & - &  g \mu_B H \sum_i (S_{i,1}^z+ S_{i,2}^z)
\label{eq-ham}
\end{eqnarray}
As for the isolated dimer, the ${\bf D}$ vector is assumed to lie in
the $yz$ plane, i.e.,
${\bf D} = (0, D \sin \theta, D \cos \theta)$.
Our choice of an alternating ${\bf D}$ vector from one rung to the
other (see Fig.~\ref{fig:model_ladder}) is motivated by symmetry
considerations.
Indeed, in a canonical ladder, the middle of each rung is an inversion
center, and the DM interaction vanishes by symmetry.  A simple way to
allow for the DM interaction to become finite without modifying the
symmetry of the exchange couplings is to assume that some buckling is
present along the ladder, as sketched in Fig.~\ref{fig:model_zigzag}.
In that case, the only mirror plane that contains a bond is the $xz$
plane, and a DM interaction with a ${\bf D}$ vector parallel to $y$ is
allowed by symmetry.  But, in this geometry, the presence of a $C_2$
axis (see Fig.~\ref{fig:model_zigzag}) implies that the ${\bf D}$
vector alternates from one rung to the other. The buckling realized in
Cu$_{2}$(C$_{5}$H$_{12}$N$_{2}$)$_{2}$Cl$_{4}$ is slightly more subtle
(successive rungs are connected by an inversion symmetry in the middle
of a plaquette), but this symmetry also implies alternating ${\bf D}$
vectors.  Note, however, that other ways of breaking the inversion
symmetry of the rungs can lead to other arrangements of ${\bf D}$
vectors.

Another
motivation to work with alternating ${\bf D}$ vectors is to
keep the perturbation
caused by the inter-dimer coupling as small as possible.  In that
respect, this choice is natural.  Indeed, as we have seen in the
previous section, the presence of a ${\bf D}$ vector on a rung induces
a staggered magnetization.  If the ${\bf D}$ vectors of neighboring
rungs $i$ and $i+1$ are equal, the moments $\langle S_{i,1} \rangle$
and $\langle S_{i+1,1} \rangle$ will also be equal, which is in
conflict with antiferromagnetic inter-rung exchange interactions.  If,
on the contrary, the ${\bf D}$ vectors are opposite on neighboring
rungs, the local moments will adopt configurations that are compatible
with the exchange.

\subsection{Symmetry analysis}

With this choice of staggered ${\bf D}$ vectors, the model possesses
the following symmetries: i) a $yz$ mirror plane going through the
middle of the rungs; ii) an inversion center in the center of the
plaquette formed by two consecutive rungs; iii) even translation
symmetries along $y$.  As long as these symmetries are not broken, the
following relations between the expectation values of local spin
operators on two neighboring rungs are expected to be satisfied:

\begin{eqnarray}
\langle S_{i,1}^{x}\rangle & = &  -\langle S_{i,2}^{x}\rangle  =   -\langle S_{i+1,1}^{x}\rangle  =
\langle S_{i+1,2}^{x}\rangle \nonumber \\
\langle S_{i,1}^{y}\rangle & = & \langle S_{i,2}^{y}\rangle  = \langle S_{i+1,1}^{y}\rangle =
\langle S_{i+1,2}^{y}\rangle \nonumber \\
\langle S_{i,1}^{z}\rangle & = & \langle S_{i,2}^{z}\rangle  =  \langle S_{i+1,1}^{z}\rangle  =
\langle S_{i+1,2}^{z}\rangle \; .
\label{eq-sym-ladder}
\end{eqnarray}
We thus define the staggered and uniform magnetizations per site as
\begin{eqnarray}
{\bf m}_s &=& (1/N) \sum_i (-1)^i \left(\langle {\bf S}_{i,1} \rangle -\langle {\bf S}_{i,2}\rangle\right)
\nonumber \\
{\bf m}_u & =& (1/N) \sum_i \left(\langle {\bf S}_{i,1}\rangle +
\langle {\bf S}_{i,2}\rangle\right) \; ,
\end{eqnarray}
where $N$ is the total number of sites, and with the convention that
the angle $\theta$ is positive for $i$ even.
As in the isolated dimer case, the staggered magnetization ${\bf m}_s$
is along the $x$ axis, while the uniform magnetization ${\bf m}_u$
lies in the $yz$ plane.

\begin{figure}
\includegraphics[width=\columnwidth]{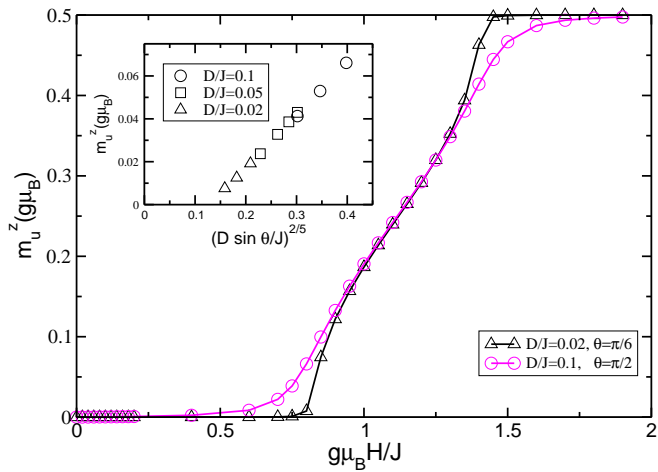}
\caption{Examples of the variation of the uniform magnetization along
  the $z$ axis with the field.
  Inset: Plot of $m_u^z$ as a function of $(D\sin \theta/J)^{2/5}$
  slightly below $H_{c1}$, which confirms the scaling predicted in
  Ref.~[\protect{\onlinecite{Fouet04}}].}
\label{fig-magz}
\end{figure}
\begin{figure}
\includegraphics[width=\columnwidth]{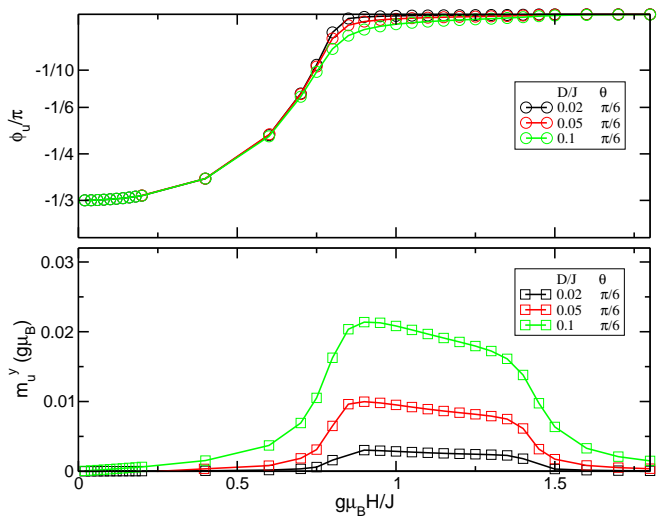}
\caption{Lower panel: $y$ component of the uniform magnetization as a
  function of the field for $\theta=\pi/6$ and various values of
  $D/J$.  Significant values appear between $H_{c1}$ and $H_{c2}$, and
  far outside this interval as soon as $D/J$ is not too small.  Upper
  panel: Angle $\phi_u$ between the magnetic field and the uniform
  magnetization ${\bf m}_u$ as a function of the field for
  $\theta=\pi/6$ and several values of $D/J$.  Note that
  $\theta-\phi_u$ goes to $\pi/2$ in the low-field limit, in agreement
  with the prediction for an isolated dimer
  (Eq.~\protect{\ref{eq-m-Hsmall}}).}
\label{fig-phi-ang30}
\end{figure}

\subsection{Uniform and staggered magnetizations}

Let us now turn to the discussion of the numerical results we have
obtained for the model of Eq.~(\ref{eq-ham}).  We are interested in the
regime $D <J_\parallel< J$.  For $D=0$, the model is a simple ladder
in a field, and the properties are well understood.  There is, of
course, no staggered magnetization because of the U(1) symmetry, and
the uniform magnetization is parallel to the field for the same
reason.  It vanishes below a critical field $H_{c1}$, takes off with a
square-root singularity, and reaches saturation with another
square-root singularity at a second critical field $H_{c2}$.  The
difference $H_{c2}-H_{c1}$ scales with $J_\parallel$.  Since, apart
from this scaling, the properties depend very little on $J_\parallel$,
we quote results for a single value of $J_\parallel$, and having in
mind the compound Cu(Hp)Cl,\cite{Chaboussant98} we have chosen
$J_\parallel/J=0.2$.  For that ratio, the critical fields in the
absence of DM interactions are given by $g\mu_BH_{c1}=0.82\,J$ and
$g\mu_BH_{c2}=1.40\,J$.

For the model with DM interaction, we have
performed Exact Diagonalization (ED)\cite{noack05} up to 20 sites (10
rungs), and DMRG calculations on ladders with up to 80 rungs.  The
results evolve smoothly with the size, and we only quote DMRG results
obtained for 80-rung clusters (finite-size effects for the gap are
discussed in the next section).
Note that in those calculations, $S_z$ is not a good quantum
number. This is well known to reduce greatly the maximal size available
to exact diagonalizations, but this also has an impact on the number of states
we were able to keep during the DMRG runs. Here, we diagonalize (by the
Davidson method) a matrix of size $4m^2$ at each DMRG step. In a
standard DMRG run where $S_z$ is a good quantum number, the matrix of
the effective Hamiltonian in the variational basis is block-diagonal,
which can speed up the diagonalization by a factor of 10 or more. The
memory needed is also larger at fixed $m$ than for the standard DMRG. For
those reasons, most of the calculations were done with up to $m=600$
states kept during 5 sweeps, and only up to $N=80$ sites. The discarded weight
was of the order of $10^{-10}$ when we targeted two states to extract
the gap, and of the order of $10^{-12}$ or less when we targeted a
single state to extract correlations. 
We also performed a few runs
with $m$ up to 800 in order to confirm that the numerical data were
well converged. 

\begin{figure}
\includegraphics[width=\columnwidth]{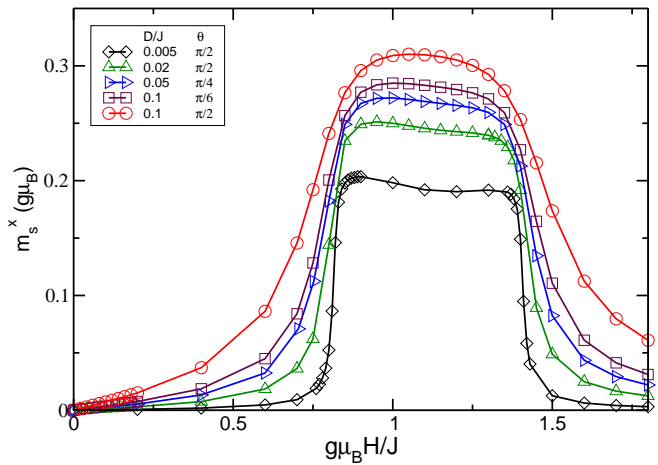}
\caption{Staggered magnetization as a function of the magnetic field
  for several values of $D/J$ and $\theta$.  Large values are achieved
  between $H_{c1}$ and $H_{c2}$, and far outside this interval as soon
  as $D/J$ is not too small.  The value between $H_{c1}$ and $H_{c2}$
  depends relatively weakly on $D/J$ and $\theta$, and is of the same
  order as the maximal value in the case of an isolated dimer
  ($0.35\,g\mu_B$).  In contrast, the value outside this interval
  depends very strongly on the magnitude of $D\sin\theta$.}
\label{fig-stmagx}
\end{figure}

The $z$ component of the magnetization is displayed in
Fig.~\ref{fig-magz} for several values of $D$ and $\theta$.  It is
reminiscent of that for $D=0$; however, when $D\ne 0$, the magnetization
develops as soon as the magnetic field is switched on, only reaching
saturation asymptotically in the limit of infinite field.  The square
root singularities are removed.  It was shown in
Ref.[\onlinecite{Fouet06}] that, at $H_{c1}$, the magnetization should
depend on the magnitude of the $D$ vector as $m_u^z \propto (D\sin
\theta)^{2/5}$, in agreement with the present results 
(see the inset of Fig.~\ref{fig-magz}).

When $\theta\neq \pi/2$ (i.e., $D_z\neq 0$), a uniform magnetization
along the $y$ axis also develops, as in the isolated dimer case.
Fig.~\ref{fig-phi-ang30} shows the magnetization along the $y$ axis
and the angle $\phi_u$ between the uniform magnetization and the $z$
axis as a function of the magnetic field for $\theta=\pi/6$.  At low
field, the uniform magnetization is orthogonal to the DM vector, again
as for an isolated dimer.  The magnetization along $y$ is maximal
between the two critical fields.  Its value in that range is clearly
much smaller than the component along the field ($\phi_u$ becomes very
small near $H_{c1}$), but this extra contribution to the uniform
magnetization will produce a torque that should be detectable
experimentally given the very high sensitivity of torque measurements.

The staggered magnetization along $x$ exhibits a kind of plateau in
the intermediate phase between $H_{c_1}$ and $H_{c_2}$
(Fig.~\ref{fig-stmagx}).  Its magnitude inside the plateau is of the
order of the maximal value of the isolated dimer ($0.35\,g\mu_B$), and
it depends relatively weakly on $D$.  In contrast, the extent of the
tails outside this plateau region increases rapidly with $D$.
Remarkably, the magnetization per spin along $x$ is larger than along
$z$ up to $H_{c1}$ and even slightly above.  Note that the staggered
magnetization depends essentially on the value of $D_y$ and is very
weakly affected by the value of $D_z$.

\vskip.5cm
\begin{figure}[bht]
\begin{center}
\includegraphics[width=\columnwidth]{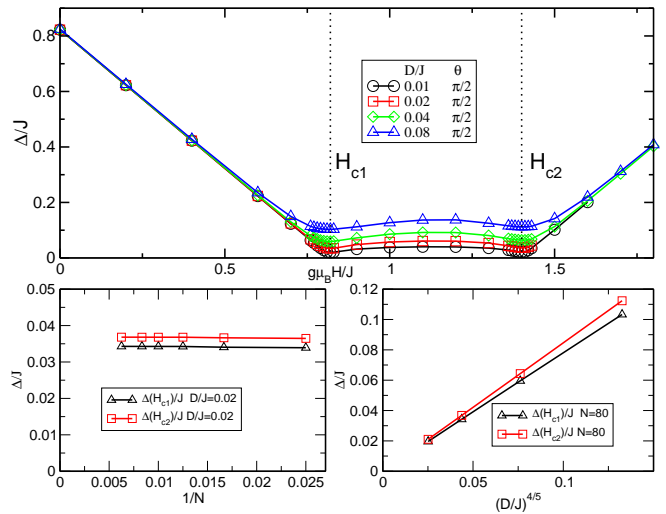}
\end{center}
\caption{Upper panel: Field dependence of the excitation gap $\Delta$ for $J_{\|}/J
  = 0.2$ and several values of $D/J$: $D/J = 0.01$(black), $D/J =
  0.02$(red) $D/J = 0.04$(blue) $D/J = 0.08$(green) and $N = 80$
  (DMRG).
Lower left panel: Scaling of the excitation gap as a function of 1/N for D/J=0.02.
Lower right panel: Scaling of the gap as a function of D/J for N=80 (see text).}
\label{fig:gap}
\end{figure}

\subsection{Gap}

The effect of a $SU(2)$ breaking interaction on a ladder has been
studied in Ref.~ [\onlinecite{Fouet06}].  It strongly depends on the
nature of the plateau phase.  For the transition from the zero or full
polarization to the gapless phase, the effective field theory is
expected to be the same as for the spin chain close to saturation, and
the gaps at $H_{c1}$ and $H_{c2}$ should open as
$(D\sin{\theta})^{4/5}$, as shown in Ref.~ [\onlinecite{Fouet04}].
This prediction clearly agrees with the results for $\theta=\pi/2$
shown in Fig.~\ref{fig:gap} (lower right panel). Size effects are already very small for
$N=80$ sites, as can be seen in Fig.~\ref{fig:gap} (lower left panel).  Between
the two critical fields, the gap is expected to remain finite.  (The
closing of the gap in Ref.[\onlinecite{Fouet06}] was caused by a
breaking of the $Z_2$ symmetry which does not occur here as there is no $m=1/2$ plateau when $D/J=0$).  The effect
of the $z$ component of $D$ is expected to be very small.  This is
also confirmed by our DMRG results (not shown).
\subsection{Torque measurement}

\begin{figure}[bht]
\begin{center}
\includegraphics[width=\columnwidth]{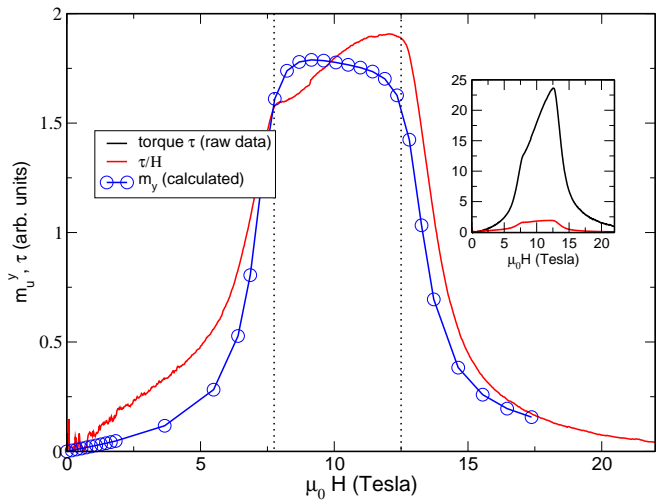}
\end{center}
\caption{(Color online) Transverse uniform magnetization $m_y$ 
for $D_y/J=0.05, D_z/J=0.086,J_{||}/J=0.2$
(blue circles) and torque divided by field (experimental curve obtained on $\rm Cu(Hp)Cl$) as a
function of the field. Inset: torque measurement raw data (black) and
torque divided by field (red).}
\label{fig:torque}
\end{figure}

If the uniform magnetization is not parallel to the field, it
induces a torque $\tau$ on the system.  Usually, such a torque
is only present if the field is not along a high symmetry
direction of the $g$-tensor. This is the basis of torque
measurements of the magnetization. However, as shown above, a DM
interaction can also induce a component of the magnetization
perpendicular to the magnetic field, which should show up in
torque experiments as an additional contribution. Interestingly
enough, torque measurements on Cu(Hp)Cl indeed reveal the presence
of such a contribution. Experiments were carried in a resistive
magnet and $\tau$ was measured up to 23~T at 410 mK.   The
orientation of the crystal was adjusted so that $\tau = 0$ at the
highest values of $H$, as shown in the inset of Fig.~\ref{fig:torque},
which fully cancels the contribution due to the 
anisotropy of the $g$ tensor. This orientation indeed corresponds
to $H_\|$ [100]. In spite of this, a large additional
contribution shows up between the critical fields, and extends
well outside the intermediate region. For comparison, the calculated
component of the uniform magnetization perpendicular to the field
of a ladder with $J_{\|}/J=0.2, D_y/J=0.05$ and $D_z/J=0.086$ 
($\theta=\pi/6$) is depicted on the same plot, with scales adjusted to 
get the same
value at $H_{c1}$. The values of $D_y$ and $J_{\|}$ are those used
in Ref.~\onlinecite{Clemancey06} to fit the staggered
magnetization, while the results depend very little on $D_z$ up to
an overall scale factor. The two curves are in good qualitative
agreement, especially considering the fact that the only
adjustable parameter is the overall scale factor.
In order to go beyond this qualitative agreement, it would be
necessary to consider several additional effects.
First of all, inelastic neutron scattering data
have challenged the description of this system as a simple ladder,\cite{Stone02}
and further couplings (still note definitely identified) should presumably
be included. 
In addition, there is a transition into a 3D ordered phase\cite{Chaboussant98}
between $H_{c1}$ and  $H_{c2}$, and although the precise nature
of the ordering is still unknown, it is very likely it will affect the uniform 
magnetization. Clearly, at the present stage, too little is known about these
additional effects to be able to take them into account, and this is left for
future investigation.

\section{Conclusions}
If spin 1/2 dimers are coupled in such a way that there is no
inversion center at the middle of the bond, very significant
modifications of the physics in a magnetic field have to be expected.
Indeed, unless it is forbidden by symmetry, a DM interaction will
always be present, and the analysis reported in this paper shows that
even a tiny DM interaction can modify some aspects of the physics
rather dramatically.  This is especially true for the staggered
magnetization, which immediately acquires large values in the
intermediate phase where the system gets polarized, and which can
take on significant values outside this phase for physically
relevant values of the DM interaction.  This is also true for the
uniform magnetization as soon as the ${\bf D}$ vector of the DM
interaction and the field are neither parallel nor perpendicular.  In
that case, a component of the uniform magnetization perpendicular to
the magnetic field appears, which can induce a measurable torque on
the sample.  This has been proven for an isolated dimer and for a
ladder with staggered DM interactions, but these conclusions are
expected to hold true for all coupled-dimer systems as long as the
${\bf D}$ vectors are arranged in such a way that there is no
competition with Heisenberg exchange as far as the development of a
staggered magnetization is concerned.  It is our hope that these
results will help understand some of the strange properties observed
in coupled-dimer systems.

\begin{acknowledgments}
We acknowledge useful discussions with Karlo Penc and Oleg
Tchernyshyov.  This work was supported by the Grant-in-Aids for
Scientific Research on Priority Areas "Invention of Anomalous Quantum
Materials" and for Aoyama Gakuin University 21st COE Program from the
Ministry of Education, Culture, Sports, Science and Technology of
Japan, by the Swiss National Fund, by MaNEP, and by the SFB 382.
\end{acknowledgments}

\end{document}